\documentclass[aps,prl,twocolumn,english,showpacs]{revtex4-1}
\usepackage[utf8]{inputenc}
\usepackage[T1]{fontenc}
\usepackage[english, french]{babel} 
\usepackage{amsmath}
\usepackage{amsfonts}
\usepackage{graphicx}
\usepackage{mathtools} 
\usepackage{braket} 
\usepackage[usenames,dvipsnames]{xcolor} 

\begin{document}
\title{Entropy driven formation of complex crystals in soft nanoparticle systems}

\author{ Anuradha Jagannathan }
\affiliation{Laboratoire de Physique des Solides, CNRS-UMR 8502, Universit\'e
Paris-Sud, 91405 Orsay, France }

\selectlanguage{english}

\date{\today}


\begin{abstract}
This paper proposes a description of a theoretical mechanism for self assembly in binary soft nanoparticle systems of the type studied experimentally by Talapin et al \cite{talapin2009}. A phenomenological free energy is written based on n-body depletion potentials, and based on this a variety of periodic as well as quasiperiodic structures are predicted to form, depending on the size ratio of the particles. We argue that this theory can qualitatively explain many of the experimentally seen structures, including striped, tetragonal, hexagonal and quasiperiodic phases, and predict some new square triangle tilings. The stability of these phases, in particular, of quasicrystals is argued to be enhanced by soft shells of the nanoparticles. This phenomenological theory could be tested by detailed numerical investigations of the depletion forces in binary systems.
\end{abstract}
\maketitle

{\subsection{Introduction}}
In a paper that has given rise to much interest, Talapin et al \cite{talapin2009} showed that a suspension of soft nanoparticles of two different sizes can self-assemble to form a variety of complex crystalline or quasicrystalline structures. In this and later works \cite{recent} the nanoparticles, consisting of a metallic core surrounded by a deformable corona, self-assemble forming different layered structures, depending on the size ratio. In addition to the hexagonal or tetragonal symmetry, a quasicrystalline structure having 12-fold (dodecagonal) symmetry was observed. Phases with 12-fold symmetry in fact are a robust phenomenon, seen not only in several binary particle systems, but also in block copolymer blends, silica mesophases and dendrimer systems \cite{ungarzeng}. The diversity of experimental systems where dodecagonal symmetry is found seems to indicate that there must exist rather general grounds for this type of symmetry to be preferred. In the systems we consider, interactions are repulsive and short ranged, indicating that the stabilizing factor is probably entropic in origin. We consider a Landau-Ginzburg approach for this type of freezing problem, in which many-body terms in the free energy arise due to so-called ``depletion forces" of entropic origin. It has been known since the work of Asakura and Oosawa \cite{asakura} that when hard spheres are immersed in a sea of smaller hard spheres, an effective attractive short-range force is developed between the former due to the latter. In this paper we put forward the idea that generalizations of this type of depletion forces, which depending on the size ratio of the spheres $q$, contribute to certain many body terms in the free energy, and these lead to the long range ordering, in particular, to quasicrystals.

The question of formation of quasiperiodic structures has been actively addressed since the discovery of quasicrystals by Schechtman et al \cite{schechtman}. In an early work, Mermin and Troian \cite{mermin} generalized a freezing model introduced by Alexander and McTague (AM) \cite{alexmctague} by making a two-length scale hypothesis. In this approach the quasicrystal is stabilized by a competition between two incommensurate distances. For 2D systems, such two-length scale mean field theories have been studied in a number of works \cite{barkan,dotera,barkanengel}.  Molecular dynamics calculations using the two-scale potentials proposed by Barkan et al \cite{barkan} confirm that different varieties of cluster crystals can self-assemble, including decagonal and dodecagonal cluster crystals \cite{barkanengel}.  Density functional calculations using two length-scale interactions have been carried out \cite{archerpapers}  reaching similar conclusions.  Numerical studies using Monte Carlo or molecular dynamics have been carried out, in particular, for hard-core-soft-shoulder (HCSS) models, where the potential is assumed to have a hard core repulsive part for the metallic cores of the particles, and a softer potential representing the repulsive interaction between the coronas of the particles. A complex phase diagram including 12-fold and higher order symmetric phases is found as the relative interaction radii and the particle density \cite{doterahcss} are varied. 

In all of these models, quasiperiodicity in the models is ``built in" from the start, via the two length scale hypothesis. In this paper, we propose a mechanism whereby complex structural phases such as large unit cell crystals, and quasicrystals, proceeds via an initial stage of cluster formation due to depletion forces followed by freezing of clusters.  In this approach, two length scale physics is an emergent phenomenon rather than imposed in the theory from the start. The length scales depend on the local clusters which preferentially form for specific particle size ratios. We begin by considering hard spheres, and argue that geometrical packing factors determine the strength and sign of the $n$-body depletion interactions. These give rise  to a variety of structures including 1D ``striped", hexagonal, square and 2D quasicrystals belonging to a class of random square-triangle tilings \cite{oxborrow}. It should be noted that, in the previously mentioned 2D theoretical models, stable quasiperiodic order, when observed, occurs in an extremely narrow range of parameters. However, as we mentioned earlier, quasiperiodic order appears in quite diverse experimental contexts, and seems to be more readily formed than these models would indicate. We argue that the formation of quasicrystals is facilitated by soft nanoparticles, where the deformable shells can lead to better stability. 

While the particles and clusters are three dimensional, we focus here on the self assembled phases which occur at an air-liquid interface and are quasi two-dimensional. The simplest cluster geometries have 2D analogs that have been enumerated in a study of a 2D binary hard disk system by Likos and Henley \cite{likoshenley}.  By direct calculation of packing fractions, they obtained an ``infinite pressure" phase diagram as a function of the composition (ratio of large and small disks) and the disk-size ratio.  It is interesting to ask what happens when particles are spheres -- rather than strictly 2D disks -- confined at an interface. A recent study investigated a somewhat easier problem, by introducing non-additive interactions in 2D to mimic three dimensionality \cite{etienne}. The results is that the Likos-Henley infinite pressure phase diagram is firstly somewhat modified, with the appearance of some new crystal structures. Furthermore, interestingly, the range of stability for certain structures such as the quasicrystalline phase, is enhanced upon introducing non-additivity.

\begin{figure}[!ht]
\centering
\includegraphics[width=150pt]{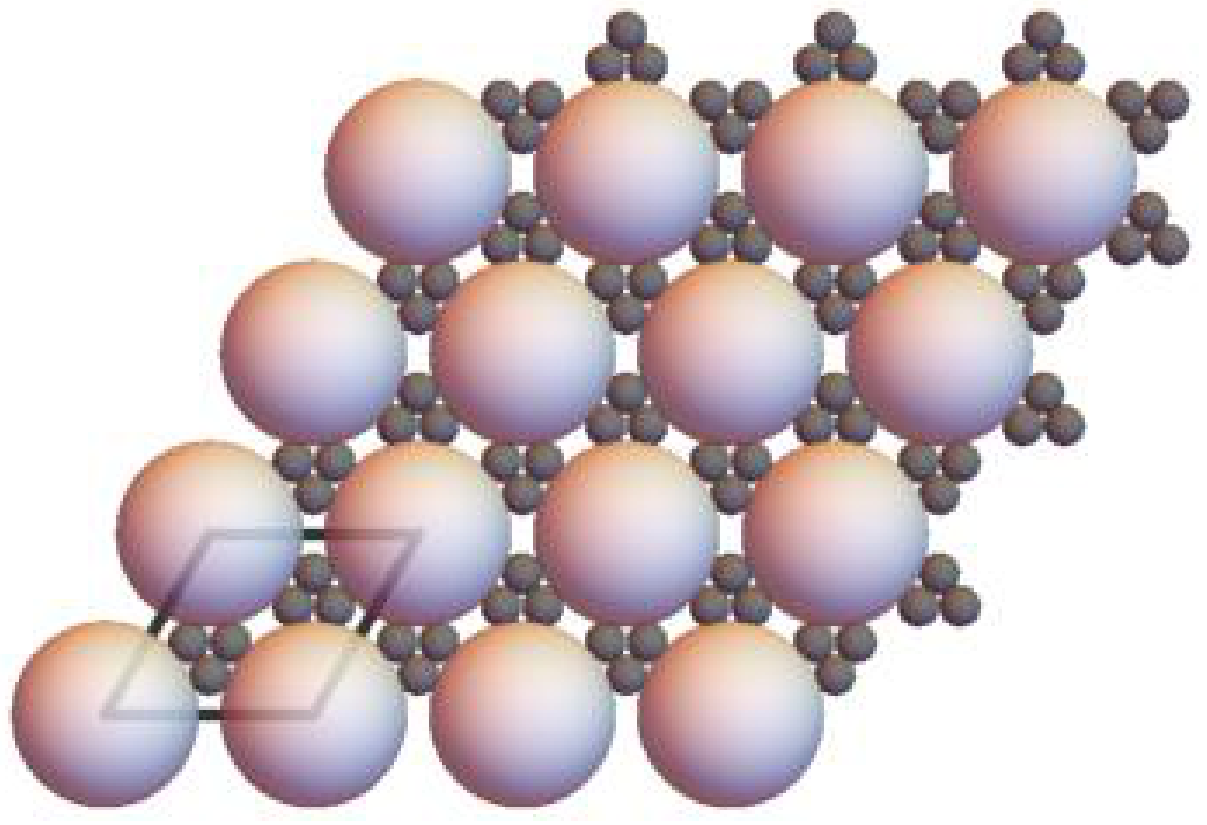} \includegraphics[width=120pt]{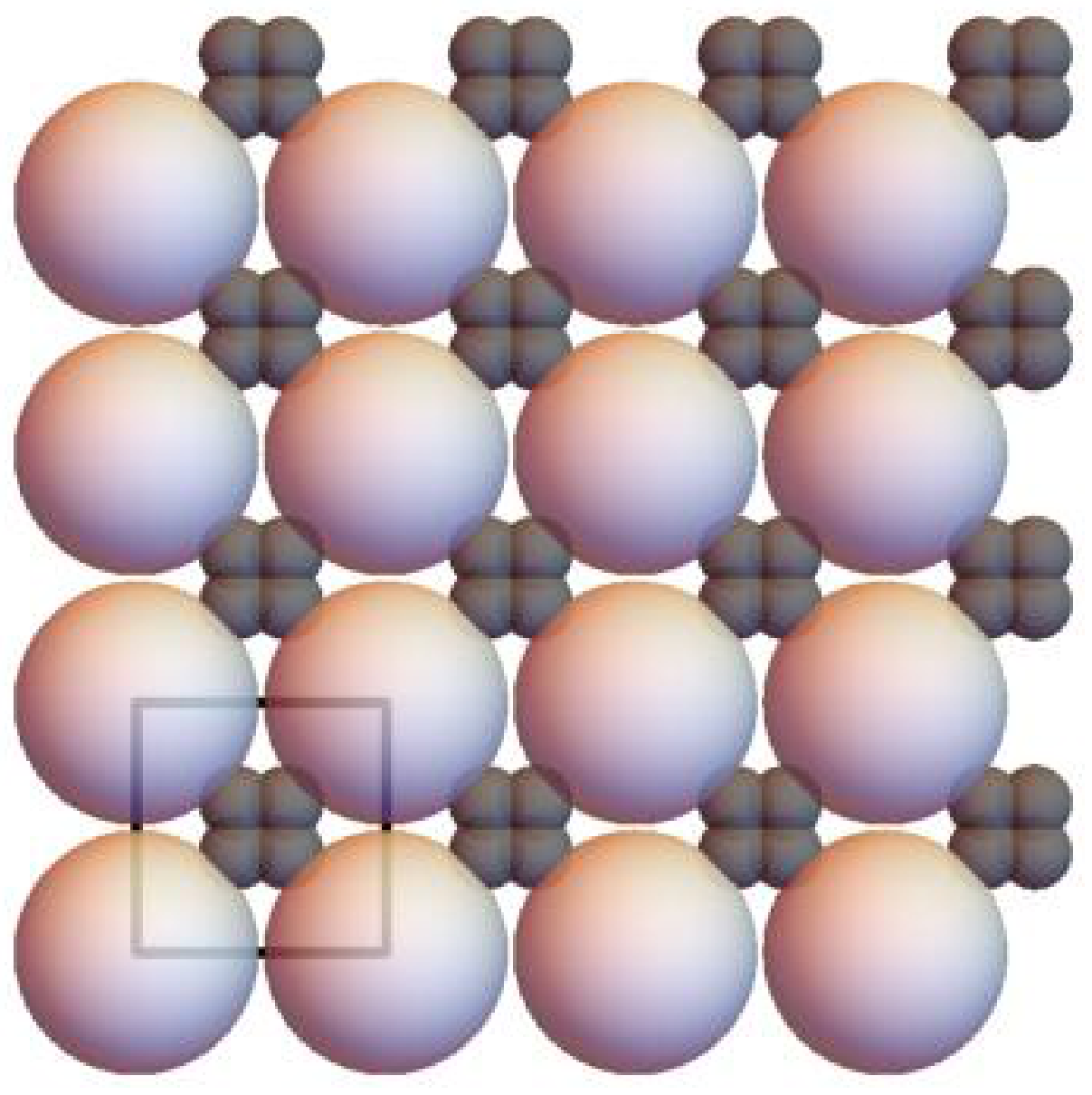}  \includegraphics[width=120pt]{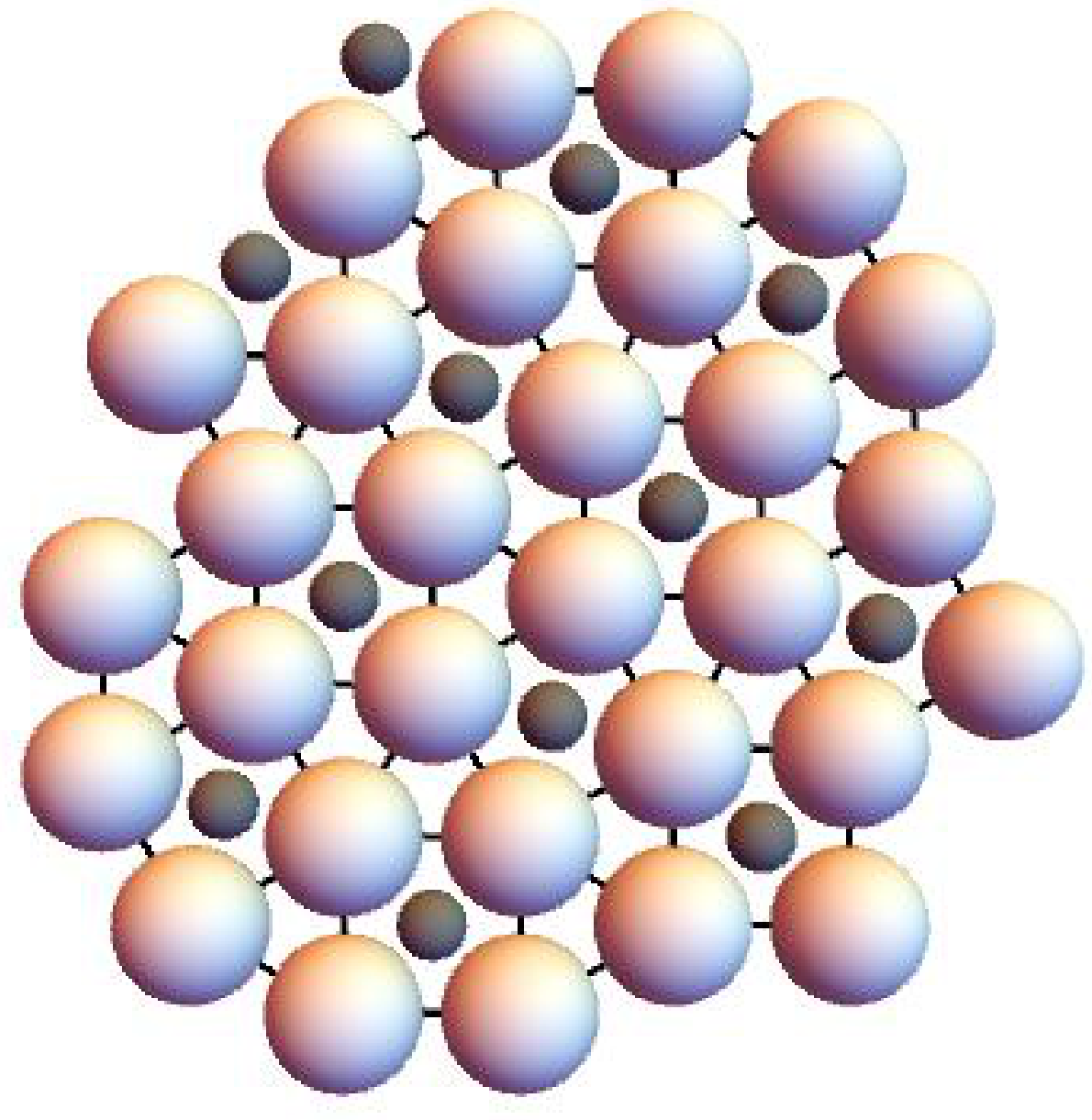}  \includegraphics[width=130pt]{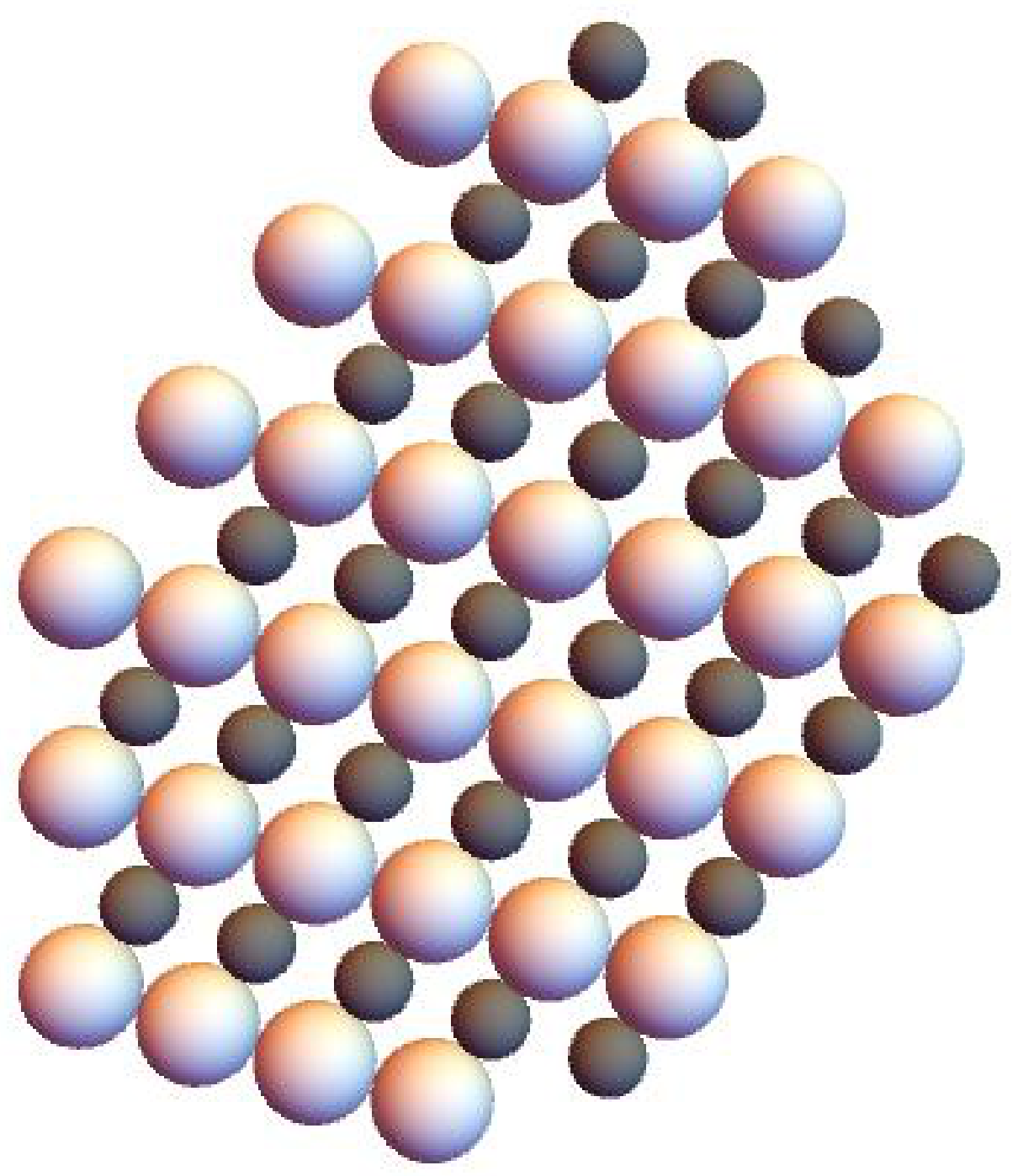} 
\caption{a) Examples of crystal structures with a) hexagonal symmetry b) tetragonal symmetry c) a square triangle tiling d) a striped phase}
\label{lattices.fig}
\end{figure}

\subsection{Description of the system and model} We consider the freezing transition in binary systems composed of large (L) and soft (S) particles, having the size ratio $\sigma=r_s/r_l$.   For concreteness we may consider these particles to be metallic nanoparticles as in the experiments of Talapin et al \cite{talapin2009}, however our approach is applicable to many analogous systems. Recall that these metallic nanoparticles have a hard core and have thiols or other long organic molecules attached to their surface, forming a soft corona around the core. The purpose of the corona is to stabilise the particles when they are in an emulsion. The resulting pair interaction potential is a sum of two contributions, $V=V_{hs}+V_{ss}$. $V_{ss}$ is the soft shell potential for interparticle distances such that only the coronas overlap and $V_{hs}$ gives the hard sphere interaction between the metallic cores at shorter distances. 

The grand partition function of the binary system is
\begin{eqnarray}
\Omega= \sum_{N_L,N_S} e^{\beta(\mu_l N_L + \mu_s N_S)} \left(\frac{ \lambda_l^3}{N_L !} \frac{  \lambda_s^3}{N_S !}\right) {\mathrm{ Tr}} e^{-\beta V}
\label{partfn.eq}
\end{eqnarray}
where $\beta=1/k_BT$ is the inverse temperature and $\mu_{l(s)}$ denotes the chemical potentials of each type of particle. The trace ``Tr" refers to the configuration integrals for the two different species $\nu=L,S$, which remain to be evaluated after the trivial momentum integrals have been carried out, resulting in the thermal de Broglie factors $\lambda_{l(s)}$.  One can then express the free energy $\beta F= -\log \Omega$ of the binary system as a sum of an ideal gas part $F_{id}$ given by 
\begin{eqnarray}
\beta F_{id}[\rho_L,\rho_S]  = \sum_{j=1,2} \int d\vec{r} ~\{ \rho_j (\vec{r})\log (\lambda_j^3\rho_j(\vec{r})) -1\}  
\end{eqnarray}
and the remaining part $F_{ex}$, the excess free energy due to interactions. In density functional theory (DFT) the free energy can be written in terms of the density fluctuations. In a single component system, for example, it is a function \cite{yrtheory} of the dimensionless density field $n(\vec{r}) =  (\rho(\vec{r}) -\overline{\rho})/\overline{\rho}$ where $\overline{\rho}$ is the average of the the spatially varying density $\rho(\vec{r})$.

The order parameters in the solid phase are $n_{\vec{q}}$ for $\vec{q}$ corresponding to the reciprocal lattice of the solid phase. To describe freezing transitions which are second order or weakly first order, it is further assumed that it is sufficient to keep the lowest order terms in the expansion of $F[n]$:
\begin{eqnarray}
\Delta F = F[n] - F_0 =  \frac{1}{2} \int d\vec{q} ~S^{-1}(q)n_{\vec{q}}n_{-\vec{q}} \nonumber \\
+ u_3  \prod_{j=1}^3 \int d\vec{q}_j ~n_{\vec{q}_1}n_{\vec{q}_2}n_{\vec{q}_3} ~\delta(\sum_j \vec{q}_j) \nonumber \\
+ u_4  \prod_{j=1}^4 \int d\vec{q}_j ~n_{\vec{q}_1}n_{\vec{q}_2}n_{\vec{q}_3}n_{\vec{q}_4} ~ \delta(\sum_j \vec{q}_j) \nonumber \\
+ u_5 \prod_{j=1}^5 \int d\vec{q}_j ~n_{\vec{q}_1}.....n_{\vec{q}_5} ~ \delta(\sum_j \vec{q}_j) + O(n^6)
\label{freeexp.eq}
\end{eqnarray}
where the free energy is measured with respect to a reference free energy $F_0$, $S(q)$ is the structure factor of the liquid, and the prefactors $u_j$ depend on the temperature and average density. 

\subsubsection{ Freezing transition and crystal symmetries}
In their insightful work \cite{alexmctague} Alexander and McTague showed that rather general arguments can be made to explain the occurrence of periodic structures of certain symmetry, according to the structure of the free energy in Eq.\ref{freeexp.eq}. They simplified the generic expression for the free energy functional, by taking the modulus of the wave-vector fixed to the value $q_m$, where the main peak of the structure factor diverges as the transition temperature is approached. In the solid phase, the set of reciprocal lattice vectors of this fixed length, is $\vec{q}_j$, with $j=1,...,M$. Here the number $M$ corresponds to the rotational symmetry of the lattice, for example, $M=6$ (hexagonal), $M=4$ (tetragonal) and $M=12$ (dodecagonal). AM assume that the coefficient $u_4$ to be positive, and that 5th and higher order terms can be neglected close to the transition. Fixing the normalization $\sum_{j=1}^M n_{\vec{q_j}}^2=1$, one gets
 
\begin{eqnarray}
\Delta f[n] = \frac{1}{2} r(T) n^2 +  u_3 c_M n^3 + u_4c'_M n^4
\end{eqnarray}
where $r(T)= a(T-T_1)$ is proportional to the inverse correlation length squared. The symmetry of the lattice is encoded in the combinatorial factors $c_M$ and $c'_M$, which can be readily found for each type of lattice \cite{chaikin}. In particular, $c_3$ is the number of triangles one can form from three reciprocal lattice vectors i.e. $\vec{q}_1 +\vec{q}_2 +\vec{q}_3 =0$ which is $2M$ for the hexagonal and dodecagonal lattices, and 0 for the tetragonal lattice. The fourth order term can be similarly evaluated. The (first-order) transition temperatures  for each $M$, $T_c^{(M)}$ can thus be determined, and in two dimensions, the winner turns out to be the hexagonal lattice. The reason that this lattice is chosen is linked to the fact that it is the densest, and therefore has the greatest entropy gain associated with its formation.
The AM approach was extended by Mermin and Troian to a two-length scale model \cite{mermin}, where the $S(q)$ is assumed to have two large peaks, representing two characteristic distances. The free energy in the MT theory depends on two coupled order parameters $n_1$ and $n_2$, corresponding to $q_1$ and $q_2$, and the result is a quasicrystal for an appropriate choice of the ratio, $\sigma$.  Similarly, Barkan et al were able to stabilize quasicrystalline phases in the 2D case \cite{barkan}. In this type of approach, as we have said in the introduction, two-length scales are put into the theory from the outset. 

We introduce now a different route to obtain an effective free energy for the binary NP system described by Eq.\ref{partfn.eq}. The structure factor peak corresponds here to a single length scale $\ell$ -- the typical interparticle distance, which is of the order of $2r_l$.  Binary systems have been discussed before using DFT \cite{ashcroft}, however the approach taken here is different. Our strategy will be to integrate out the small particles, and express the total free energy $F$ as a  power series of the density fluctuations, $n_L(\vec{r})$. For simplicity of notation the subscript ``L" will be suppressed henceforth. We will assume, in the derivation of the free energy, that particles are hard spheres, and that the $n$-body interaction terms are only important for systems in which stable clusters can be formed due to depletion forces, generalizations of the Asakura-Oosawa force \cite{asakura} as discussed in the next section.

For simplicity, we consider that the solid is two dimensional and forms in the $z=0$ plane, however, the arguments can  be generalized to include structures which repeat periodically along $z$). The density components $n(\vec{q})$ are henceforth defined for two-component vectors $\vec{q} = (q_x,q_y)$ by
\begin{eqnarray}
n_{\vec{q}} = \int dx\int dy ~e^{i(xq_x+yq_y)} n(x,y,0) 
\end{eqnarray}
with the corresponding inverse transform giving $n(x,y,0)$ in terms of $n_{\vec{q}}$.

\subsection{Effective n-body interactions and depletion forces}
We recall that the Asakura-Ookawa potential  describes the attractive short range force between large colloidal particles immersed in a solution of macrolecules. It describes the fact that, when large particles approach within a sufficiently small distance, there results a local expulsion of small particles, whose net entropy rises. Said equivalently, the free energy is lowered due to an effective attractive force called ``depletion force" between the L particles.  The Asakura-Ookawa attractive potential $\beta W(h) = -\frac{3}{4r_s^2} \eta_s(h-2r_s)^2 $ for $0<h<2r_s$ (where $h$ is the distance between the surfaces of the large spheres along the line joining their centers) , is only the first term of an expansion in powers of the density $\rho_S$. A discussion with more details of analytical calculations for the form of the potential along with comparison with the results from molecular dynamics simulations can be found in \cite{gotzelmann}.  Fig.1 shows a sketch of the expected form of this potential, which has a hard repulsive core, is attractive at very short range, develops a few oscillations at larger distances and decays quickly to zero. This entropically generated force has been measured in experiments, as for example in a study of micrometer sized large PMMA spheres in a solution of much smaller polystyrene spheres \cite{crocker}. 
These considerations on the depletion forces in a pure dilute system must be modified in our binary problem, which has several important differences. Firstly, the size ratios $\sigma$ of interest to us, ranging from close to zero to around 0.5, are not small so that the simple approach using the Derjaguin approximation is not applicable. Secondly, the medium in which the two L particles are placed is a two-component mixture, so that the pressure and surface tension are more complicated to compute as compared to the systems in \cite{gotzelmann}. Finally the particles in experiments are not perfect hard spheres but deformable. The behavior of the effective depletion-type forces in generalizations of the simple Asakura-Oosawa binary system have been reported in the literature \cite{biben,cinacchi,binder}. These find that the Asakura-Oosawa form holds for small size ratio $q \leq 0.15$ or so. Larger size ratios lead to the generation of 3- and 4-body terms \cite{santos}. These are precisely the terms that are important for our theory. 

Neglecting the deformability of the particles, we now consider colloidal solutions of hard L and S spheres, for which the depletion pair potential is attractive at short range, as in Fig.1.  

\begin{figure}[!ht]
\centering
\includegraphics[width=150pt]{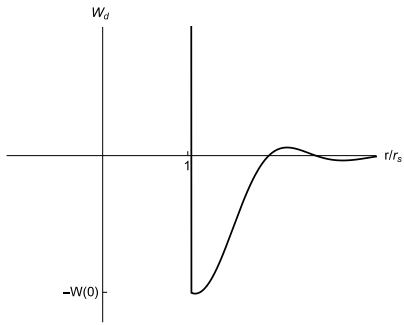} 
\caption{Sketch of the expected form of total potential as a function of the distance (measured in units of $r_s$) from the center of a large particle, showing the repulsive hard core and the attractive region. }
\label{fullpot.fig}
\end{figure}

The structure factor $S$ of the L particles is found perturbatively by using the relation between $S$ and the Fourier transform of the direct correlation function (DCF), $\hat{c}(q)$, $S^{-1}(q) =(1-\overline{\rho} \hat{c}(q))$. The DCF can be expressed in terms of the radial distribution function $g$ and the pair potential function $\phi$ through $c(r)=g(r)[1-e^{\beta \phi}]$, where the qualitative form of the potential $\phi$ is shown in  Fig.\ref{fullpot.fig}. Note that we do not need an exact functional form for our present purposes but only the main salient properties namely, the hard core extends to a distance $r_l$, and for larger $r$, there is an attractive part followed by an oscillating part out to a distance of roughly $2r_s$.
 In the absence of the depletion potential, the pair potential is the hard sphere potential $\phi=\phi_{hs}$, for which the forms of the radial distribution function, $g_{hs}(r)$ and DCF, $c_{hs}(r)$ have been computed (see \cite{hansen}). The change of DCF, $\delta c(r)$, in the presence of the depletion potential can be estimated as follows
\begin{eqnarray}
\delta c(r) \approx -\beta g_{hs}(r) W(r) 
\end{eqnarray}
where we have made an expansion to linear order in $W$ assuming that the depletion potential is a small perturbation and used $e^{\beta \phi_{hs}}=1$ for $r\geq 2r_l$. Replacing the exact form of the depletion potential by $W(r)= W(0)r_s \delta(r-2r_l)$ allows to write the 2D Fourier transform of $\delta\hat{c}(k)$ explicitly as
\begin{eqnarray}
\delta\hat{c}(k)=- 2\pi \beta g(2r_l) W(0) J_0(k)
\end{eqnarray}
 where $J_0$ is the zero order Bessel function of the first kind. When added to the hard sphere DCF, the depletion interaction results in small modifications of the peak heights and positions of this function. For our purposes the structure factor can be written in the simple form \cite{chaikin}
\begin{eqnarray}
\beta S^{-1}(q) = a(T-T_1) + c_2(q^2-q_m^2)^2
\end{eqnarray}
where  $T_1 = \overline{\rho}\hat{c}_m$, the spinodal temperature, is proportional to the value of $\hat{c}$ for $q=q_m$. The constants $a$, $q_m$ and $c_2>0$ can be determined from fits to the structure factor, and the values are expected to be slightly modified from those of the single component hard sphere system, due to the depletion force. 

In dense systems, higher order terms in $\rho_L$ play an important role. In fact as we will see, the third and fourth terms play the most important role in determining the symmetry of the solid phase. Introducing the quantities $\epsilon_n$ for the strengths of the $n$-body terms ($n>2$), the free energy expansion is given by Eq.\ref{freeexp.eq} with
\begin{eqnarray}
u_3&=&\left(\frac{-T}{6}+\epsilon_3(\sigma, \rho_s,\rho_l)\right)  \nonumber \\
u_4&=&  \left(\frac{T}{12}+\epsilon_4(\sigma,\rho_s,\rho_l)\right)  \nonumber \\
u_5&=&  \left(\frac{-T}{20}+\epsilon_5(\sigma,\rho_s,\rho_l)\right)
\label{amfree.eq}
\end{eqnarray}
where the $\epsilon_n$ represent n-body depletion interactions, and depend strongly on the size ratio $\sigma$ and the volume fractions of the particles. Importantly, and in contrast to the two body potential, these $n>2$-body interactions are not expected to be smooth functions of the size ratio, because the binary system has special geometrical packing properties for ``magic" values of $\sigma$, as listed by Likos and Henley \cite{likoshenley} in their study of a 2D binary system of hard disks at $T=0$.  A detailed analytical and numerical calculation of the values of these interactions as a function of $\sigma$, $\rho_l$ and $\rho_s$ is beyond the scope of this paper.  Our focus here rather is on showing the different symmetries of phases which can be expected to form for different choices of the parameters $\epsilon_n$. The key observations are 1) that $\epsilon_n$ is largest (most negative) when $n$ large particles form a close packed cluster, where the depletion forces can play a role and 2) to contribute to the free energy,  the stoichiometry of the system should allow for formation of a macroscopic number of these $n$ clusters. The former condition leads to the strong dependence of $\epsilon_n$ on size ratio, $\sigma$  and the latter is determined by the packing fraction $p$. 

We conclude this discussion with remarks on the role of the soft shells surrounding particles. Results for depletion forces for deformable particles  have been considered \cite{cinacchi}, where it was shown that even a slight increase of softness resulted in a somewhat surprising increase of the attractive region of the depletion force. For $n$ body interactions, more generally, the effects of particle deformability should lead in the first approximation to lowering the corresponding free energies. This is because the overall cluster volume is lowered due to the ability of deformable particles to fill the voids space more efficiently. This could particularly favor aperiodic phases, where a larger number of voids shapes is found, as compared to the simpler periodic phases.

In the next sections we show that, depending on the strength and sign of the $n$-body depletion interactions, Eq.\ref{freeexp.eq} can lead to a variety of phases, including 1D ``striped" phases, hexagonal, square and quasiperiodic phases, depending on the signs and magnitudes of $u_3$ and $u_4$. The structures which are actually found would also depend on the packing fraction of the mix. As in the preceding literature in the field, this list of phases is non-exhaustive -- the possibility that we missed some as yet-unknown favorable structure cannot be ruled out. The existence of a soft shell is particular interesting in the case of Case 2, the quasicrystal.

\subsubsection{A. Hexagonal lattices} While the depletion force aided formation of triangles of large particles is favorable, such that $u_3$ is large in magnitude, but the 4-body depletion potential $\epsilon_4$ is small enough so that the coefficient $u_4$ in Eq.\ref{freeexp.eq} is positive, our model is equivalent to the AM model. A hexagonal phase is therefore to be expected. In experiment \cite{talapin2009} , a lattice of the AlB$_2$ type has been seen -- in this structure each triangle of L particles has an S particle in the interior. The maximum size ratio corresponding to fitting the small particle in the void is the ``magic" size ratio $\sigma_1 = 0.16$ as noted by \cite{likoshenley}. Another interesting possible triangular structure, corresponding to a smaller size ratio $\sigma$, is illustrated in Fig.\ref{lattices.fig}a) for an $AB_6$ system (where A and B stand for the L particles and S particles respectively). This particular lattice is called the T3 phase in \cite{likoshenley}, but many other compositions can exist in the form of 6-fold symmetric cluster phase. Hexagonal lattices are the most commonly occurring phase in the Likos and Henley phase diagram of 2D hard disks. They constitute a major part of our phase diagram as well. However, 12-fold symmetry can be more favorable than hexagonal symmetry for a large region of parameter space, as shown in the next section.

\begin{figure}[!ht]
\centering
\includegraphics[width=100pt]{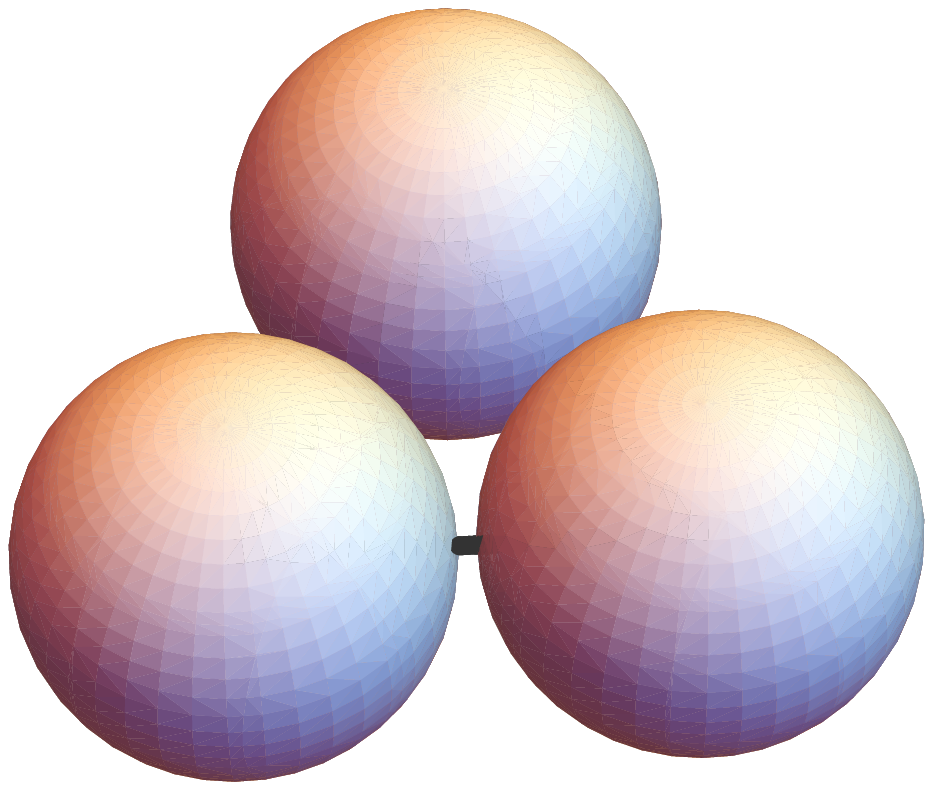} 
\includegraphics[width=100pt]{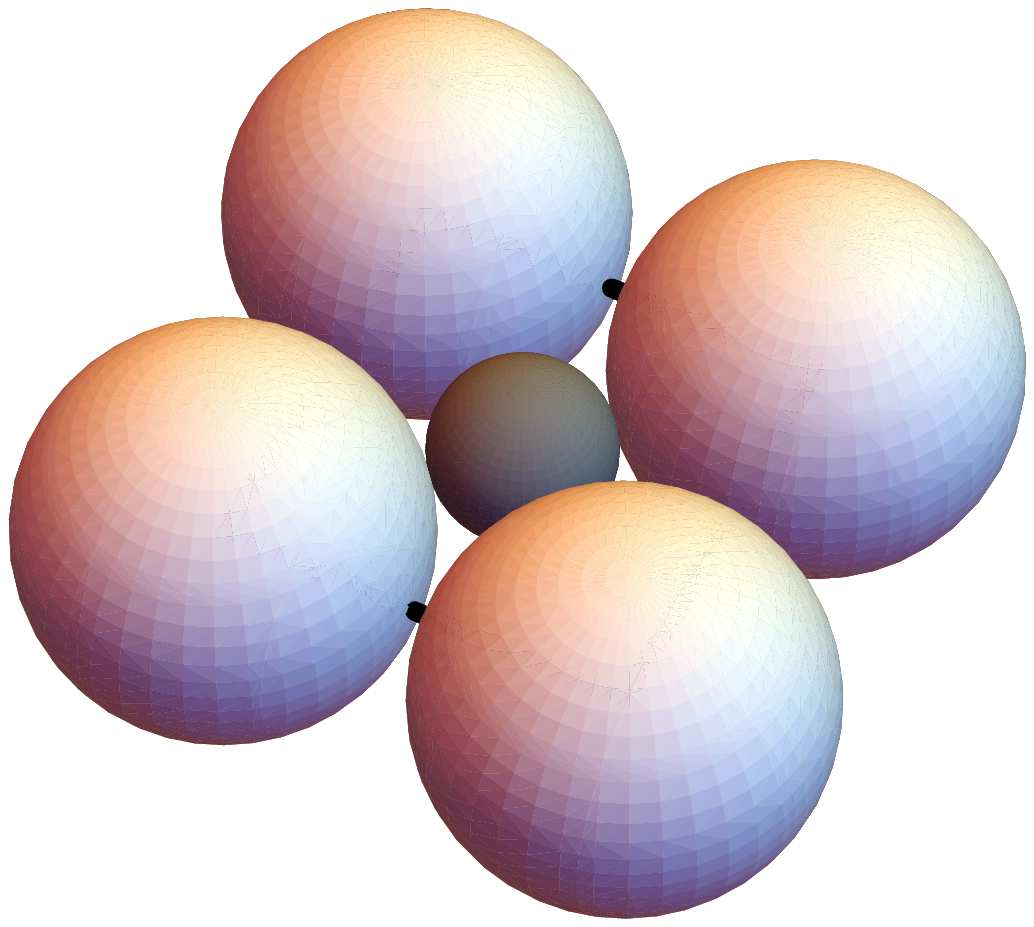}
\label{clusters1.fig}
\caption{ Compact triangular and square clusters of edge length $\sim 2r_l$ for L(light gray) and S(dark gray) particles for size ratio $\sigma=0.4$ }
\end{figure}

\subsubsection{B. Quasicrystals}  
{\bf{1. The case $\sigma \sim 0.41$}} The size ratio $\sigma_2=\sqrt{2}-1$ corresponds to one of the ``magic numbers" discussed in \cite{likoshenley}, for which a small disk fits exactly within the central vacancy formed by four large disks. In this case, depletion forces can stabilize two small clusters of the same edge length: the empty equilateral triangle and the square with one S particle inside shown in  Fig.\ref{clusters1.fig}. To give a rough estimate : $\epsilon_3 \propto 3W(0)$, where the factor 3 represents the number of L-S contact points. Similarly, the squares could be expected to have a binding energy proportional to $\epsilon_4  \propto 4W(0)$. 

\begin{figure}[!ht]
\centering
\includegraphics[width=200pt]{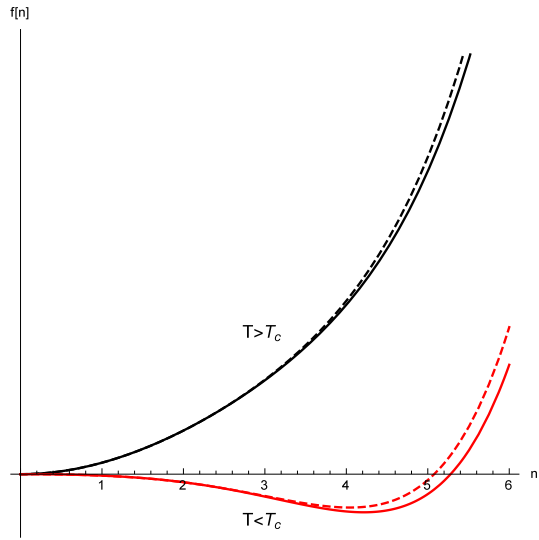} 
\label{freeplot.fig}
\caption{Plot of the free energy versus order parameter $n(M)$  as given by Eq.\ref{quasicrystalfree.eq}, for high (black) and low (red) temperatures, for the case $M=6$ (dashed lines) and $M=12$ (solid lines). }
\end{figure}

When the packing fraction $p=3-\sqrt{3}/4 \approx 0.32$, a 12-fold symmetric quasicrystal phase can occur, as was pointed out by Likos and Henley. For entropic reasons, this quasicrystal is most likely to be a random tiling composed of squares and triangles, examples of which can be found in \cite{oxborrow}. For the case of interest, namely the 12-fold symmetric case, these random tilings are composed of equilateral triangles and squares (Fig.2), in the proportion $N_\triangleleft/N_\diamond = 4/\sqrt{3}$ -- i.e. the total area covered by the triangles is equal to that covered by squares. For this set of random tilings, it is known that there is quasi-long range order, in the sense that diffraction intensities are not delta functions in the infinite size limit but decay algebraically \cite{henley}.The relative number of large and small particles is given by
\begin{eqnarray}
N_L/N_S = N_\diamond/(N_\diamond+N_\triangleleft/2) = 2\sqrt{3}-3 \approx 0.46
\end{eqnarray}
providing a condition for the chemical potentials for this structure.
Going back to the free energy expansion in Eq.\ref{freeexp.eq},  when $u_4$ is negative, it is necessary to consider the 5th order term for stability of the theory.  This free energy can be used to obtain solutions numerically. However we will draw some qualitative predictions by making simplifications along the same lines as Alexandre-McTague. All Fourier components $n_{\vec{q}}$ are neglected, excepting for components of $\vert \vec{q}\vert=q_m$, leaving a single family of $M$ order parameters, $n_{\vec{q}_1}=n_{\vec{q}_2}=...= n$, for a structure with $M$-fold rotational symmetry. The values of $M$ for 2D structures of principal interest are $M=2,4,6,12$. The resulting free energy density can be written using the AM normalization as
\begin{eqnarray}
\beta f[n] \propto \frac{a}{2\tilde{T}}(\tilde{T}-1) n^2 - \frac{u_3}{\sqrt{M} }n^3 - u_4 n^4 \nonumber \\+ \frac{u_5}{\sqrt{M}} n^5 +  u_6 n^6
\label{quasicrystalfree.eq}
\end{eqnarray}
for the two most competitive structures $M=6$ and 12, $\tilde{T}=T/T_1$ is the reduced temperature, $u_3$ and $u_4$ are positive near the transition. Fig.\ref{freeplot.fig} shows the form of $f$ (in arbitrary units) for the choice $a/2\tilde{T}=100$, $u_3=u_4=5u_5=1$ and $u_6=1/30$, for which the critical temperatures are $T_c^{(6)}=T_1+0.11$ for $M=6$ and $T_c^{(12)}=T_1+0.12$ for $M=12$. The plot shows the free energy for a temperature $T$ lying above the two critical temperatures (black curves) and a temperature lying below the critical temperatures (red). The free energy curve for $M=6$ (dashed) lies above that of $M=12$ (solid), showing that the dodecagonal symmetry is preferred in this case. 

The 12-fold symmetric square triangle tiling sample of such a square triangle network is shown in Fig.\ref{lattices.fig}c. In our simplified calculation, where the order parameter corresponds to a single infinite peak of the structure factor, the free energies of a perfect tiling and of a random tiling are equal. However, one of the main differences between the perfect and random tilings is the fact that the structure factor is broadened in the latter case, leading to a set of order parameters $n(\vec{q})$ in a range of $q$-values. This would introduce a small free energy difference between the random and the perfect  12-fold tilings. More to the point, the theoretical treatment does not address issues related to the dynamics of formation of random versus perfect tilings, and the difficulty of reaching the theoretically predicted stable configurations. It seems plausible to assume that given a random initial condition, the particles will most likely tend to freeze into a random tiling structure.

Note that we have included only a single component $n_{\vec{q}} =2\pi/\ell$, corresponding to the most probable distance $\ell$ in the structure. If further components are considered, the quasicrystal structure may be even easier to stabilize. As for quantitative estimates, the value of $T_c$ depends on various parameters, in particular on the contact value of the depletion potential, $W(0)$. These must be estimated by comparing with numerics.

\begin{figure}[!ht]
\centering
\includegraphics[width=100pt]{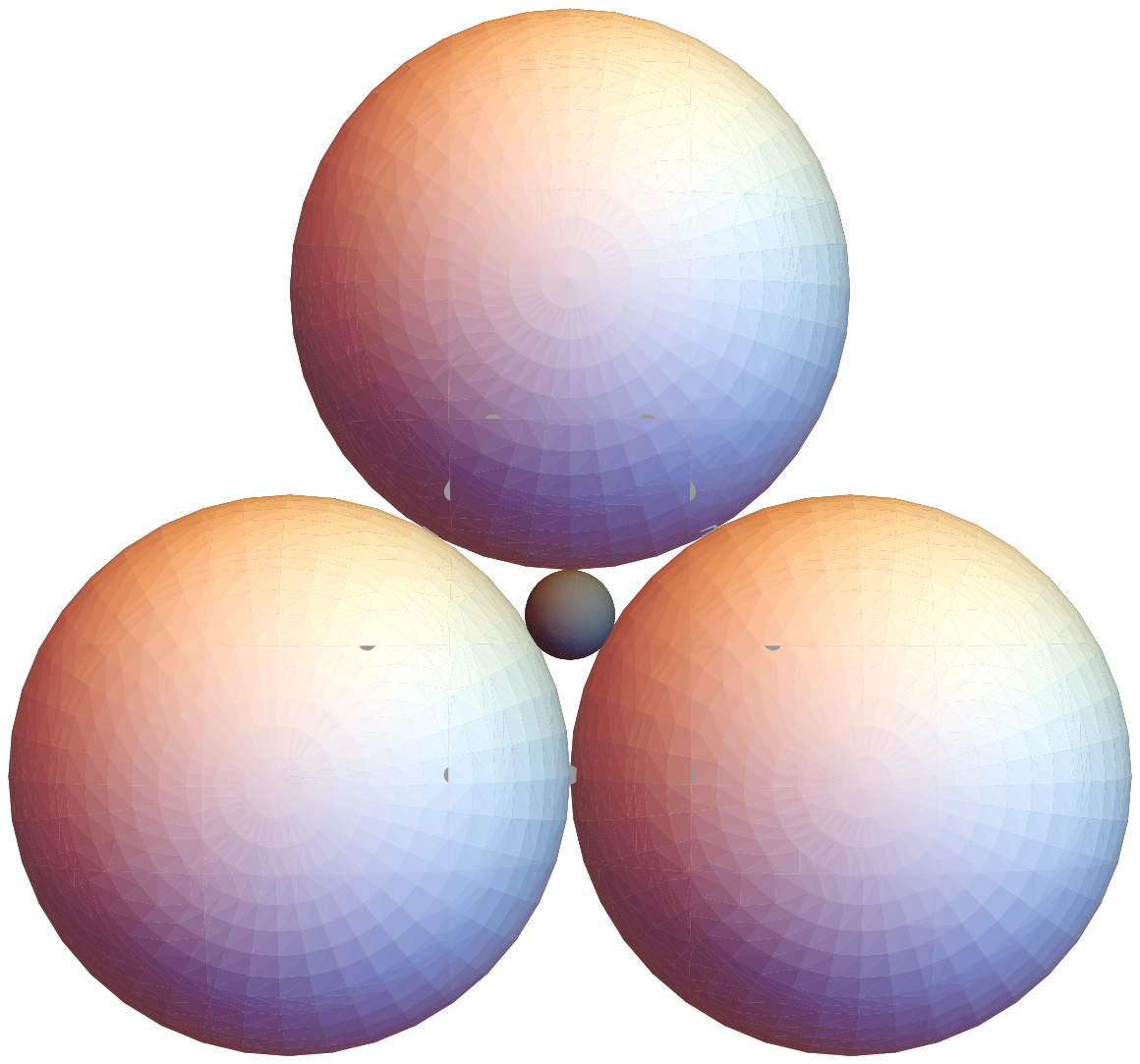} \includegraphics[width=100pt]{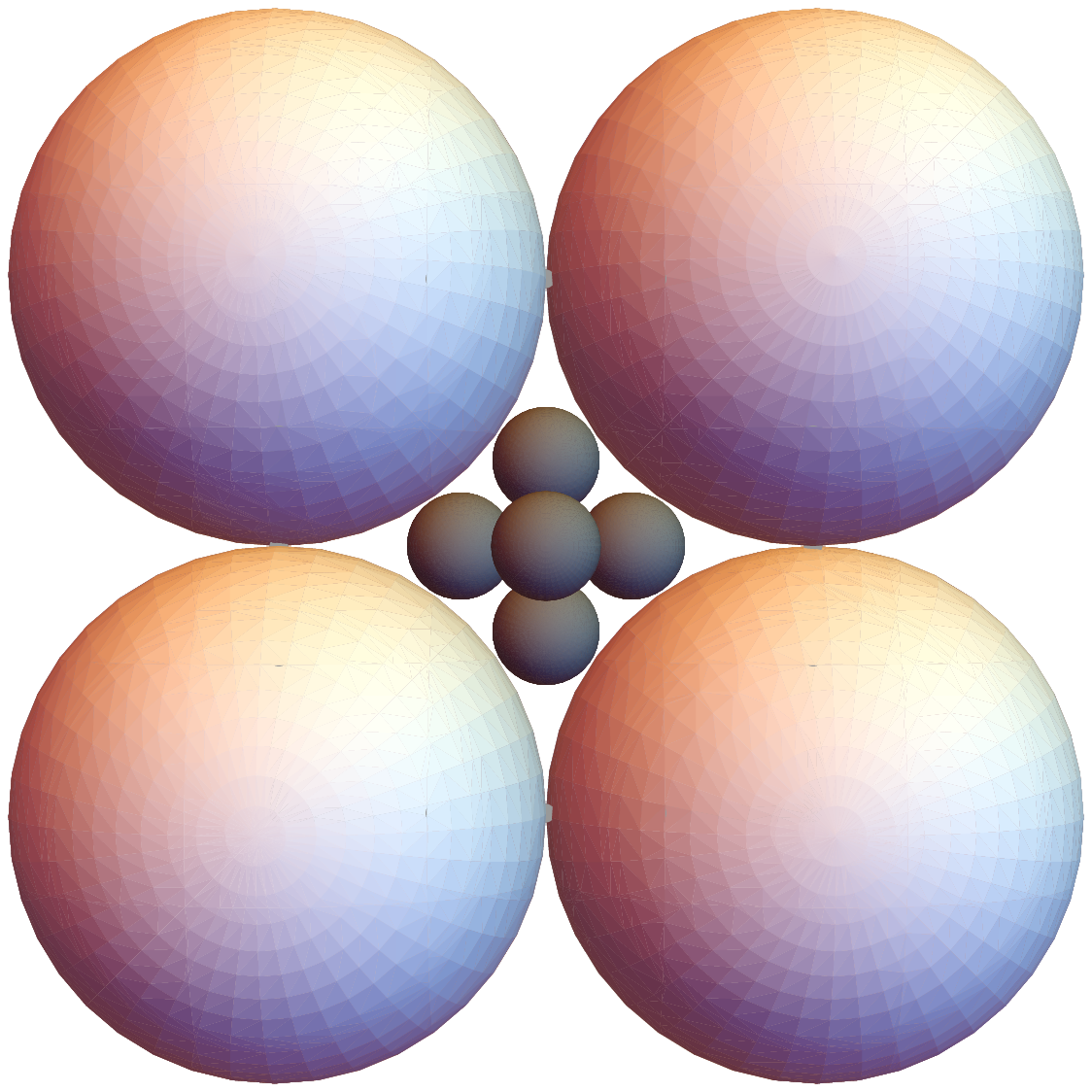}
\includegraphics[width=100pt]{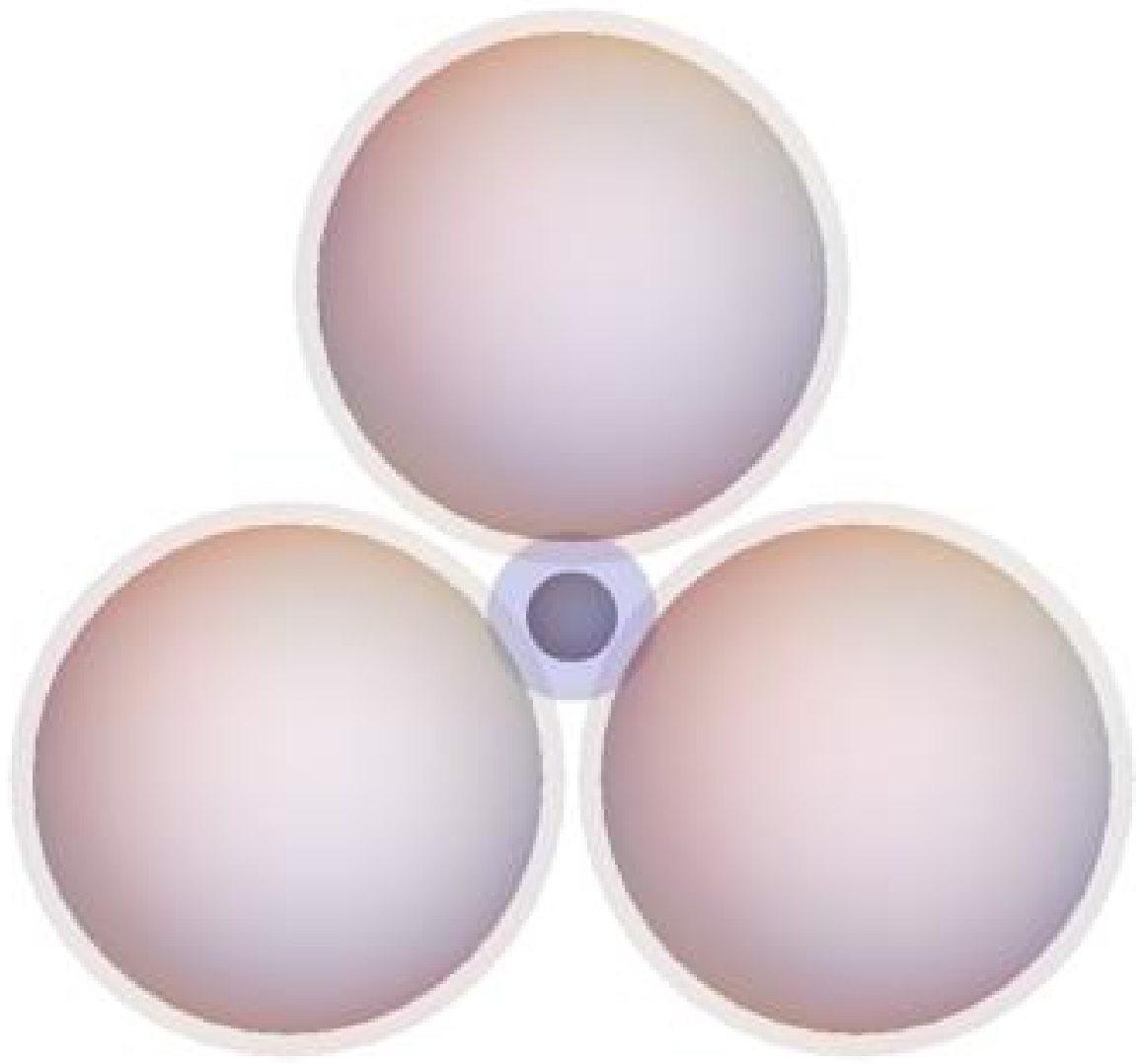} \includegraphics[width=100pt]{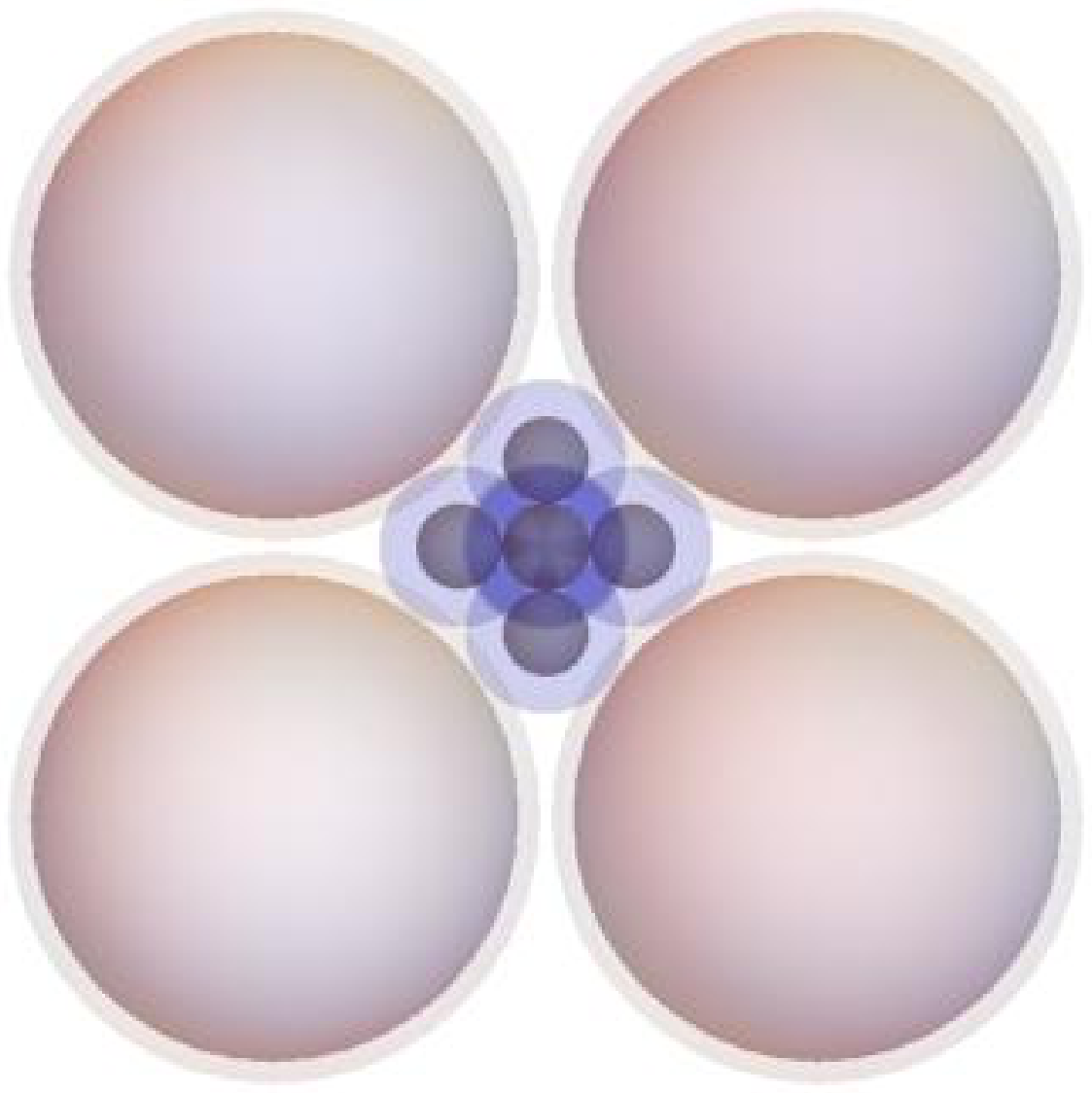}
\caption{(upper) Compact triangular and square clusters for magic size ratios $\sigma=0.16$ and $\sigma=0.22$. (lower) Same for soft nanoparticles of the same core and corona sizes.}
\label{clusters2.fig}
\end{figure}

{\bf{2. The case $\sigma \sim 0.2$ }} We will now use the fact that the particles in our binary system have soft shells, to make an argument for a new quasicrystalline phase. The shells modify the values of $W(0)$ and the $\epsilon_n$ as mentioned earlier. In addition the soft shells can play a more interesting role. When spheres are deformable, the rules of compact cluster formation are modified. This can lead to a stabilization of new  clusters forming a dodecagonal QC in a similar way as in the previous section. Let us consider two magic values, $\sigma_1$ and $\sigma_3$. $\sigma_1$ is as already noted the size ratio needed to form a compact triangle of three L particles with a single S particle in the interior. $\sigma_3 \approx 0.22$ is the size ratio needed to form a compact square of the same edge length, containing an octahedron of S particles ($\sigma\approx 0.22$).  The upper row of Fig.\ref{clusters2.fig} shows these two types of compact clusters.  For hard spheres the difference in the values of $\sigma$ make it impossible to form simultaneously both these close packed clusters. However, for our soft spheres, it should be possible to choose the hard core and soft shell thickness such that stabilization of $both$ the triangles and squares becomes possible as illustrated in the lower row of Fig.\ref{clusters2.fig}. Then, we can expect the three- and four- body terms  $\epsilon_3^{(2)}$ and $\epsilon_4^{(2)}$ to be negative and favoring the appearance of such clusters. An argument similar to the preceding case can lead to a dodecagonal quasicrystal -- which, to repeat, would not be possible for hard spheres. The fraction of small particles in the structure can be readily deduced from the known ratio $N_\triangleleft/N_\diamond$ in the quasiperiodic tiling, and is given by
\begin{eqnarray}
p =\frac{4+6\sqrt{3}}{6+7\sqrt{3}} \approx 0.79
\end{eqnarray}
The effective critical temperature can be computed for this system of particles given information on the values of $\epsilon_n$ for this case. Interestingly, note that while the first QC phase has not been observed, this second QC corresponds to precisely the type of cluster arrangement of small and large particles which is seen in the original paper of Talapin et al \cite{talapin2009}.

Finally, that other quasiperiodic phases could in principle occur for even smaller size ratios $\sigma$ by allowing small particles to fill the triangular and square voids inside the 3- and 4-body clusters. For example, triangles of L particles containing tetrahedra of S particles could become favorable for $\sigma \sim 0.1$ (in 2D this corresponds to the T2 phase in \cite{likoshenley}). A square triangle tiling could occur if squares of the same edge length can be stabilized simultaneously with an internal cluster of S particles.  The latter might, for example, be icosahedron of S particles, or, even possibly one of the new clusters seen to form in dense confined systems by Teich et al \cite{teich}, such as an octahedron made from a nine S particles ! This opens up, at least in theory, the intriguing possibility of stabilizing an infinite series of square triangle packings. However in real systems the depletion potentials per volume will presumably decrease with the number of particles, and these type of random tiling would therefore be rather unlikely to form. 

\subsubsection{C. Striped and tetragonal phases.} 
\begin{figure}[!ht]
\centering
\includegraphics[width=200pt]{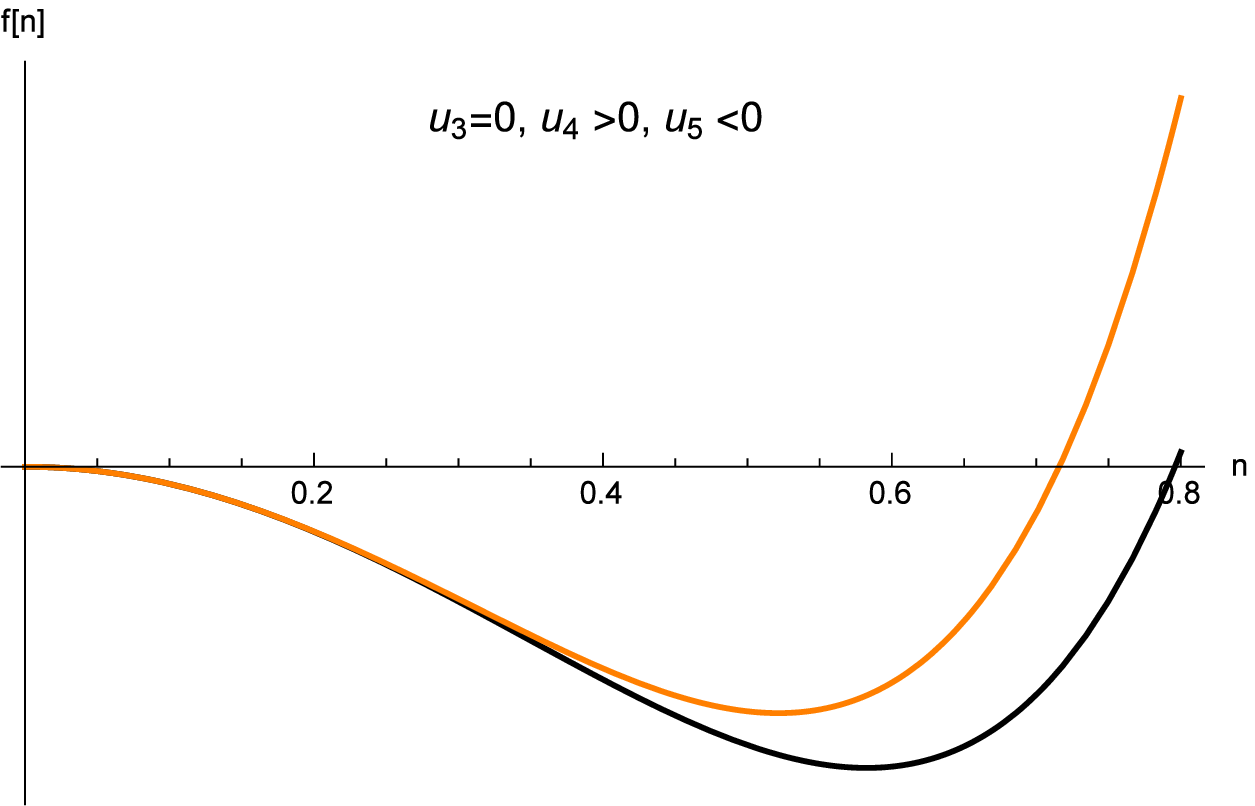} \includegraphics[width=200pt]{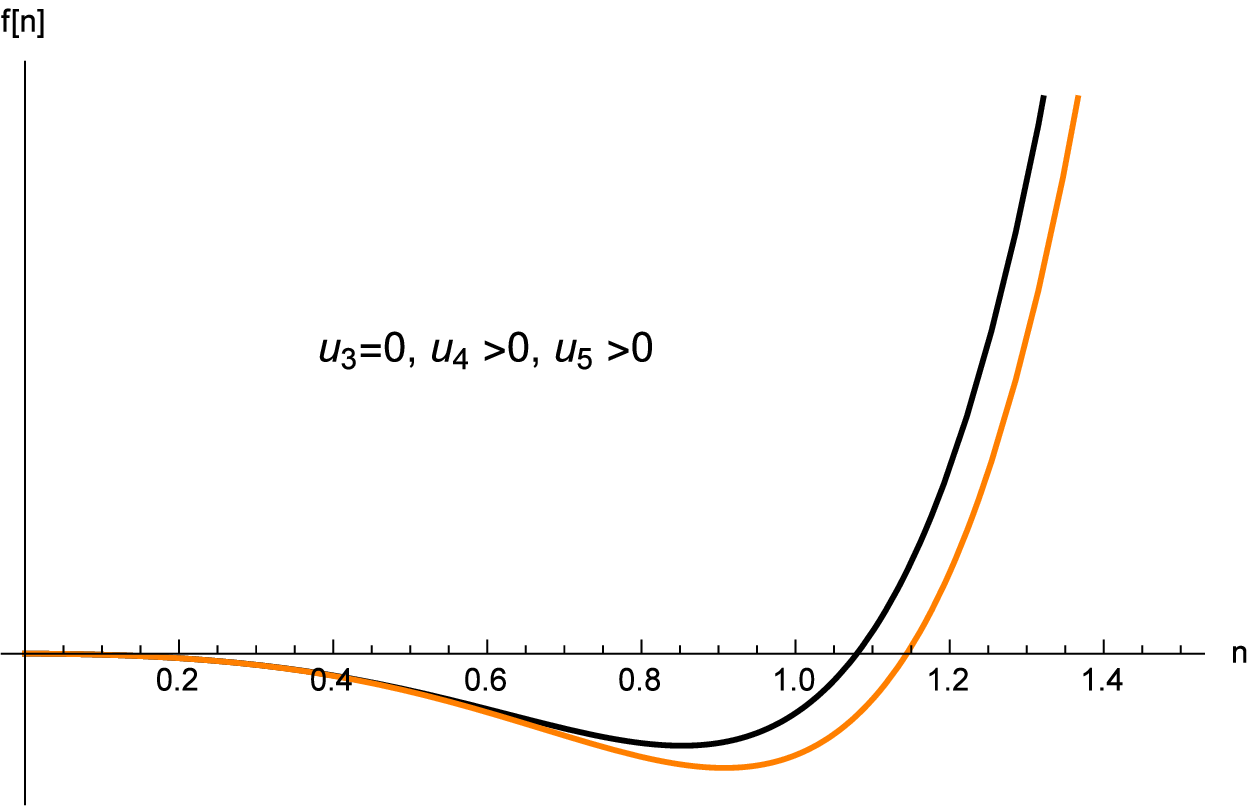}
\caption{ Plots of free energies when $u_3=0$ for $T<T_c$ as a function of order parameter $n$ for two phases: $M=2$, (stripe phase, black curve) and $M=4$ (tetragonal phase, orange curve). Parameters are such that $u_4 >0, u_5<0$ (upper figure)  and $u_4 <0, u_5> 0$ ((lower figure).}
\label{stripesquare.fig}
\end{figure}

\begin{figure}[!ht]
\centering
\includegraphics[width=170pt]{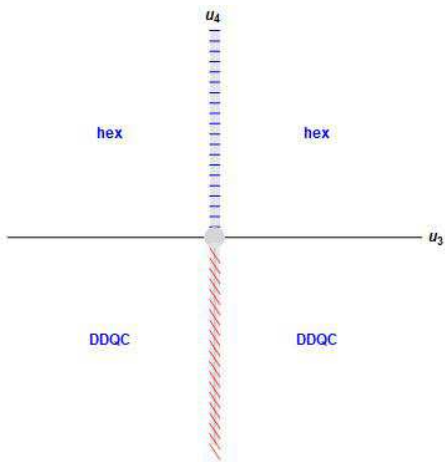}
\caption{ Phase diagram indicating the locations of stable low temperature phases discussed in the text in the $u_3-u_4$ plane. The blue striped region (for $u_3 \approx 0, u_4>0$) represents a 1D striped phase, and the red striped phase (for $u_3 \approx 0, u_4<0$ to the S2 phase. The values of the $u_5$ term vary (see text)}
\label{phases.fig}
\end{figure}

Note that the depletion potential has a repulsive part for certain distances. It could thus happen for particular size ratios and particle volume fractions that the coefficient of the third order term $u_3$ becomes very small or zero. This is shown by the hatched grey region in the phase diagram of Fig.\ref{phases.fig}. The free energy as a function of $M$ now has the form $\beta f[n]= r n^2 + u_4 n^4 + u_5 n^5/\sqrt{M} + n^6$, to sixth order in the free energy, where $r=a(T-T_1)$.  When $u_4$ is positive and $u_5$ is negative,  the comparison of free energies in the broken symmetry low temperature regime shows that the stripe phase is preferred over the tetragonal phase.  This is shown in the upper Fig.\ref{stripesquare.fig} where the free energies of the two phases are plotted as a function of the order parameter, for  the striped (M=2, orange) and tetragonal (M=4, black) phases. The temperature parameter has been chosen such that one is in the broken symmetry phase (for free energy parameters $r=-0.5$, $u_4=1$, $u_5=-1.5$).  This type of 1D striped order is illustrated in Fig.1d.  When $u_4 <0$, a lattice with square symmetry tetragonal symmetry is preferred, as seen from the lower Fig.\ref{stripesquare.fig} (for free energy parameters $r=-0.5$, $u_4=-1.5$, $u_5=1$).  A candidate for this type of stable tetragonal phase is illustrated in Fig.\ref{lattices.fig}b), in the case of $AB_4$ stoichiometry corresponding to a packing fraction $p=4/5$. Note that this structure, named S2 in \cite{likoshenley}, has been seen in experiment, where it is identified as the CaB$_6$ type superlattice \cite{talapin2009}. 

\subsubsection{D. Other phases and higher order symmetries} 
It could happen -- by accident or by design --  that for certain systems five-fold rings are favorable clusters, while three-and four-fold clusters are suppressed. This case, corresponding to negligibly small $u_3$ and $u_4$, would lead to a five-fold or ten-fold (decagonal) phase. Generalizations to higher symmetries can be made, in principle, to include for example 18-fold or 24-fold symmetries, as in \cite{doterahcss}. Such scenarios would need a careful tuning of the $u_n$ via density and size parameters and are of course highly unlikely to occur spontaneously in binary systems.

\subsection{Conclusions} We have proposed a phenomenological Ginzburg Landau model for a binary soft nanoparticle system for which depletion forces are the driving mechanism behind the freezing transition. The crystals which form are complex assemblies of clusters of large and small atoms that form due to depletion forces. We have sketched a phase diagram as a function of two of the main parameters, $u_3$ and $u_4$, and indicated the principal crystalline and quasicrystalline phases which are likely to be found. The list is indicative of possible phases at certain chosen packing fractions, and it is certainly not exhaustive. The predicted structures include striped, hexagonal, tetragonal and dodecagonal phases, depending on the size ratio and relative concentrations of the particles.  The experiments  \cite{talapin2009} indeed report seeing several of the structures, including the dodecagonal QC phase, and some hexagonal and tetragonal phases. Other new phases such as the $\sigma=\sigma_1$ quasicrystal are predicted by the theory but remain to be found in experiment. While in our simplified analysis, the size ratio $\sigma$ is the most important parameter entering in the formation of solid phase, a more quantitative analysis will require additional parameters to properly describe the the free energy. These include  more details as to the structure of the nanoparticles : relative thicknesses of the coronas and the core radii, and the characteristic energy scale of deformation of the coronas. In this paper we have argued that an important role of the soft coronas is to promote and stabilize quasicrystal formation. This is due to their deformability which relaxes geometrical constraints and allows a more efficient packing in the solid in a wider range of parameters than would be possible with hard spheres. 

As said before, our line of approach differs from the Landau-Ginzburg two-length scale hypothesis used in previous works, in that competing length scales emerge naturally as a consequence of entropy driven cluster formation. In contrast with  the hard core soft shoulder (HCSS) models \cite{doterahcss}, the symmetry and type of the preferred phase is determined principally by the size ratio of the two particles $\sigma$, while the thickness of the soft coronas should be important in promoting stability of that phase. In this context it is interesting that the quasicrystalline structure we predict for $\sigma\sim 0.2$ is seen in experiment \cite{talapin2009}, but not the simpler one predicted for $\sigma=0.4$.

In future work, it will clearly be necessary to do detailed quantitative investigations of the depletion potentials in binary systems. These will allow to check if the mechanism proposed here is in fact capable of explaining the experiments  on a quantitative level.  
Experimental work is needed to map out a systematic phase diagram, with the goal of identifying some of the new square triangle tilings predicted by our scheme. The problem of three dimensional structures, not addressed here, could be in principle be treated using our model of depletion forces. In this context, the work on three dimensional cluster formation by Teich et al \cite{teich} should provide a good starting point.

\subsection{Conflict of Interests} There are no conflicts of interest to declare.

\subsection{Acknowledgments} I would like to thank Marianne Imperor,  Jean-Fran\c{c}ois Sadoc, Etienne Fayen, Giuseppe Foffi, Frank Smallenburg and Emmanuel Trizac for useful discussions. I would like to acknowledge support from the Agence Nationale de Recherche under grant number ANR-19-OTP-62018.

\end{document}